\documentclass[aps,prl,twocolumn,showpacs,amsmath,amssymb,superscriptaddress]{revtex4}
\usepackage{graphicx,dcolumn,bm,epsfig,epstopdf,comment,color}

\def\LSCO{La$_{2-x}$Sr$_x$CoO$_4$}

\begin{document}
\title{Magnetic phase diagram of La$_{2-x}$Sr$_x$CoO$_4$ revised using muon-spin relaxation}
\author{R. C. Williams}
\author{F. Xiao}
\author{T. Lancaster}
  \affiliation{Centre for Materials Physics, Durham University, Durham, DH1 3LE, United Kingdom}
\author{R. De Renzi}
\author{G. Allodi}
\author{S. Bordignon}
  \affiliation{Dipartimento di Fisica e Scienze della Terra, Universit\`{a} degli Studi di Parma, Viale delle Scienze 7A, I-43124 Parma, Italy}
  \author{P. G. Freeman}
    \altaffiliation{Present address: Jeremiah Horrocks Institute, University of Central Lancashire, Preston PR1 2HE, United Kingdom}
  \affiliation{\'{E}cole Polytechnique F\'{e}d\'{e}rale de Lausanne, ICMP, Lab Quantum Magnetism, CH-1015 Lausanne, Switzerland}
  \affiliation{Institut Laue-Langevin, 71 avenue des Martyrs 38000 Grenoble, France}
  \author{F. L. Pratt}
\affiliation{ISIS Facility, STFC Rutherford Appleton Laboratory,
  Chilton, Didcot, Oxfordshire, OX11 0QX, United Kingdom}
\author{S. R. Giblin}
  \affiliation{School of Physics and Astronomy, Cardiff University, Queen's Buildings, The Parade, Cardiff, CF24 3AA, United Kingdom}
\author{J. S. M\"{o}ller}
\altaffiliation{Present address: Neutron Scattering and Magnetism, Laboratory for Solid State Physics, ETH Z\"{u}rich, CH-8093 Z\"{u}rich, Switzerland}
\affiliation{Oxford University Department of Physics, Clarendon Laboratory, Parks Road, Oxford, OX1 3PU, United 
Kingdom}
\author{S. J. Blundell}
\author{A. T. Boothroyd}
\author{D. Prabhakaran}
\affiliation{Oxford University Department of Physics, Clarendon Laboratory, Parks Road, Oxford, OX1 3PU, United Kingdom}
\date{\today}

\begin{abstract}
We report the results of a muon-spin relaxation ($\mu$SR) investigation of \LSCO{}, an antiferromagnetic insulating series which has been shown to support charge ordered and magnetic stripe phases and an hourglass magnetic excitation spectrum. We present a revised magnetic phase diagram, which shows that the suppression of the magnetic ordering temperature is highly sensitive to small concentrations of holes. Distinct behavior within an intermediate $x$ range ($0.2 \leq x \lesssim 0.6$) suggests that the putative stripe ordered phase extends to lower $x$ than previously thought. Further charge doping ($0.67 \leq x \leq 0.9$) prevents magnetic ordering for $T \gtrsim 1.5~{\rm K}$.
\end{abstract}
\maketitle

\begin{figure}[ht]
\centering
\includegraphics[width=7.cm]{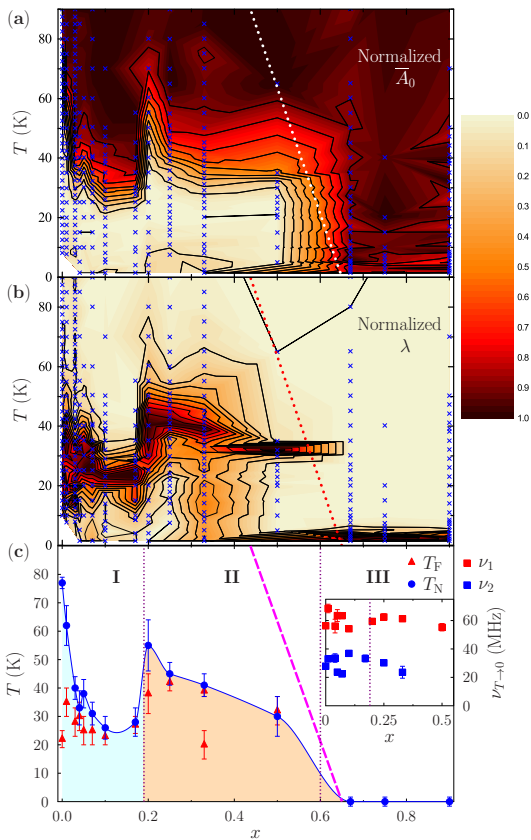}
\caption{\label{fig:heatmap} (a) Normalized early-time initial asymmetry $\overline{A}_0$ in the $x$-$T$ phase diagram for \LSCO{}. (b) Normalized relaxation rate $\lambda$. Crosses identify measured points and the dashed line shows the phase boundary from Ref. \cite{cwik}. (c) Revised phase diagram showing the nominal transition temperature $T_{\rm N}$ and freezing temperature $T_{\rm F}$. 
Inset: fitted low-temperature muon precession frequencies across the doping range $x \leq 0.5$.}
\end{figure}

Nearly 30 years after the discovery of high-$T_{\mathrm{c}}$ superconductivity, the role of multiple order parameters in the cuprates continues to be a topic of intense research, with the nature of the interplay between fluctuating stripe-like correlations of charge order (CO), antiferromagnetic spin order (SO) and superconductivity the topic of much investigation \cite{drachuck,keimer,fradkinA,fradkinB,tranquadaA,lee, yamadaA,zaanen,tranquadaB,fujita,kivelson}. 
In order to elucidate this physics, it is illuminating to study isostructural non-superconducting $3d$ transition metal oxides \cite{ulbrich} such as the antiferromagnetic series \LSCO{} (LSCO). Controlled doping of holes into the two dimensional CoO$_2$ layers is achieved via substitution of Sr for La but, in contrast to the cuprates, the series remains an insulator for $x \lesssim 1$, with electrons well localized via a spin-blockade mechanism \cite{chang}.
The previously reported phase diagram of LSCO, determined by neutron scattering \cite{cwik}, shows a region of nearest-neighbor antiferromagnetism (nnAFM) for low dopant concentrations ($x\lesssim 0.3$) and incommensurate (IC) magnetism consistent with stripe-like CO and SOs for an intermediate doping range $0.3 \lesssim x \lesssim 0.6$. Recently, muon-spin rotation ($\mu$SR) and NMR measurements  made on the stripe ordered $x= 1/3$ compound \cite{lancaster} revealed the importance of a range of timescales in the magnetic ordering of the material, 
with $\mu$SR showing both the onset of static magnetic order and the freezing of dynamical processes
at temperatures significantly lower than the ordering temperature previously identified using neutrons. 
This is because $\mu$SR is sensitive to fluctuations on the microsecond timescale (a scale set by the muon gyromagnetic ratio $\gamma_{\mu} = 2\pi \times 135.5~{\rm MHz~ T^{-1}}$). Such slow fluctuations appear static in neutron scattering measurements where the energy resolution $\Delta E \approx 1~{\rm meV}$ constrains the sensitivity of the technique to fluctuations with a much faster timescale $\hbar / \Delta E \approx 10^{-11}~{\rm s}$. 
This finding motivated this $\mu$SR investigation of the full phase diagram of the LSCO system reported in this paper, allowing us to probe the influence of slow fluctuations on the low-temperature magnetism across this series and determine a revised phase diagram. 



Crystals of LSCO were grown for our measurements using the floating-zone method, with varying sintering conditions depending on the Sr content $x$ \cite{SIfootnote}. 
Crystals grown this way have a tendency for excess oxygen when $x \lesssim 0.3$, similar to the La$_{2-x}$Sr$_{x}$NiO$_{4+\delta}$ system \cite{nemudry,girgsdies,dreau,prabhak}. On this basis we expect that the lowest doped samples have $\delta > 0$, but approach oxygen stoichiometry by $x \approx 0.3$.
Our $\mu$SR measurements \cite{blundell99,SIfootnote} were made on single crystal samples in zero applied magnetic field using the EMU spectrometer at ISIS, and the GPS and DOLLY instruments at the Swiss Muon Source (S$\mu$S), with the initial muon polarization directed along the $c$-axis of each crystallite, perpendicular to the planes containing the CoO$_2$ layers.

In data measured at S$\mu$S spontaneous damped oscillations are visible  at low temperature in the asymmetry spectra for all concentrations with $x \leq 0.5$  [Fig.~\ref{fig:heatmap}(c), inset]. 
These oscillations are indicative of quasistatic local field at the muon sites and confirm the presence of SO in each material. The two precession frequencies observed in most samples indicate two distinct classes of muon site with local field strengths in the approximate ratio 2:1 (although for certain concentrations only one frequency is resolvable in the data). There appears to be very little doping dependence in the magnitude of the two precession frequencies, supporting the claim that Co$^{3+}$ ions remain in the low-spin $S=0$ state throughout this doping range \cite{hollmann,helme}, and that there is little variation in the moment size of the magnetic $S=3/2$ Co$^{2+}$ ions.


Before considering detailed analysis, 
we provide a broad survey of the concentration-temperature $(x,T)$ phase diagram that forms the main result of this paper. 
To do this, it is illustrative to consider the average early-time asymmetry $\overline{A}_0 \equiv \langle A(t \leq 0.13~{\rm \mu s}) \rangle$ for each temperature point, where the time averaging allows us to identify the depolarization of the muon spin ensemble due to static or slowly fluctuating magnetism, shown  in Fig.~\ref{fig:heatmap}(a) (normalized between extreme values for each asymmetry spectrum \cite{footnote}). 
At high temperatures, well above $T_{\rm N}$, 
the muon spins are insensitive to rapidly fluctuating Co$^{2+}$ moments (motional narrowing), leading to large values of $\overline{A}_{0}$.
For concentrations $x \leq 0.5$ initial asymmetry is lost upon cooling, reflecting the presence of large internal magnetic fields, such as those giving rise to the oscillations described above. 
This drop in $\overline{A}_0$ may be fitted \cite{steele} with a Fermi-like broadened step function $
\overline{A}_0 (T) = A_{\rm H} + (A_{\rm L}-A_{\rm H})/(e^{(T-T_{\rm c})/w} + 1)$,
which provides a transition temperature  $T_{\mathrm{c}}$ by parametrizing the continuous step of width $w$ from high- to low-$T$ asymmetry, $A_{\rm H}$ and $A_{\rm L}$, respectively \cite{footnote2}. An onset temperature for magnetic ordering may be obtained using $T_{\rm N} = T_{\rm c} + w$ with uncertainty $w$, and these values are shown in the revised phase diagram Fig.~\ref{fig:heatmap}(c).
The previous $\mu$SR study on $x=0.33$ also demonstrated the technique's sensitivity to slow fluctuations and the freezing of dynamics \cite{lancaster}. To follow these features
 all datasets were heavily binned to filter out high frequency components and fitted to 
$A(t) = A_{\rm rel} e^{-(\lambda t)^{\beta}} + A_{\rm b},$
where $A_{\rm rel}$ and $A_{\rm b}$ are the relaxing and non-relaxing baseline amplitudes, respectively and $\beta$ was constrained to be greater than 0.5 \cite{SIfootnote}. 
The resulting relaxation rates $\lambda$  (also normalized)  are shown in Fig.~\ref{fig:heatmap}(b). 
It is expected that the relaxation rate $\lambda \propto \langle (B-\langle B \rangle)^2 \rangle \tau $ (i.e.\ the second moment of the magnetic field distribution multiplied by the correlation time $\tau$ \cite{hayano}) reflecting both field distribution widths and fluctuation rates. A peak in $\lambda$ is often indicative of a freezing of dynamics on the muon timescale as correlation times diverge, and so nominal freezing temperatures $T_{\rm F}$ (defined as the temperature corresponding to peak values of $\lambda$) feature on the phase diagram [Fig.~\ref{fig:heatmap}(c)]. 

This approach allows us to distinguish
three distinct regions in the phase diagram. For $x < 0.2$ (Region I hereafter) the introduction of non-magnetic $S=0$ Co$^{3+}$ ions suppresses the ordering temperature of the commensurate nnAFM order, and there is evidence of a freezing of dynamics at lower temperatures for compounds with small $x$. For intermediate doping concentrations ($0.2 \leq x \lesssim 0.6$, Region II) 
the behavior is consistent with that previously observed for $x=0.33$, implying that 
CO within the CoO$_2$ layers stabilizes IC stripe-like SO with a cluster glass nature. 
For $x \gtrsim 0.6$ (Region III) the system remains in the disordered paramagnetic state down to low temperatures. We examine the data in each region in more detail below. 


\begin{figure}[t]
\centering
\includegraphics[width=7.5cm]{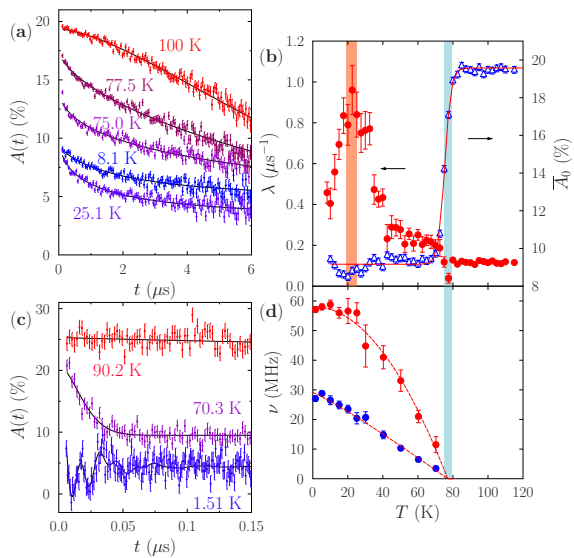}
\caption{\label{fig:00} (a) Asymmetry data for $x=0$ measured at ISIS. (b) Relaxation rate used to determine $T_{\rm F}$, plus early-time asymmetry $\overline{A}_0$ used to locate $T_{\rm N}$ [non-normalized data from Fig.~\ref{fig:heatmap} (b) and (a), respectively]. Shaded regions show the ordering and freezing temperatures, $T_{\rm N}$ and $T_{\rm F}$, respectively. (c) Data measured at S$\mu$S, showing spontaneous precession at low temperature (1.51~K data shifted by $-5\%$ for clarity). (d) Temperature evolution of the two precession frequencies.}
\end{figure}

{\it Region I:} 
Data measured on the $x=0$ compound La$_{2}$CoO$_4$ at S$\mu$S show spontaneous oscillations in the muon polarization for temperatures below $T \approx 75~{\rm K}$ [Fig.~\ref{fig:00}(c)], confirming a transition to SO. The asymmetry spectra within the ordered regime were found to be best fitted for $t \leq 0.5~{\rm \mu s}$ to the two-frequency relaxation function
$A(t) = \sum\nolimits_{i=1,2} A_i\cos(2 \pi \nu_i t) {\rm e}^{- \lambda_i t} + A_{\rm rel}{\rm e}^{- \lambda_{\rm rel} t},$
where the high and low frequency oscillatory components, with amplitudes $A_1$ and $A_2$, are  fixed to their average values of 6\% and 10\%, respectively, and indicate two magnetically inequivalent muon stopping sites. The fitted values of the two precession frequencies $\nu_{1,2}$ are plotted in Fig.~\ref{fig:00}(d), and drop to zero at a temperature $T_{\rm N}$, as expected for a magnetic transition to a disordered, paramagnetic state.
Measurements made at ISIS are well suited to probing the slow dynamics of the system [Fig.~\ref{fig:00}(a)]. An abrupt drop in average early-time  asymmetry is apparent upon cooling below $T \approx 75~{\rm K}$ [Fig.~\ref{fig:00}(b)] and a fit to the broadened step function yielded an ordering temperature value of 
$T_{\rm N} = 77(2)~{\rm K}$, consistent with the vanishing of the oscillations.  
Upon further cooling below $T \approx 70~{\rm K}$ $\overline{A}_0$ does not change, but the longitudinal relaxation rate increases to a peak at around 22.5~K, before dropping again at lower temperatures. This is suggestive of a freezing of dynamics (as the correlation time $\tau$ increases) at much lower temperatures than the transition to magnetic long range SO. This feature is shared by the $x=0.01$ sample, which could indicate that the introduction of frustration by small concentrations of $S=0$ Co$^{3+}$ ions induces a freezing transition for a relaxation channel that has a characteristic timescale within the $\mu$SR dynamical range. The freezing is visible in the asymmetry spectra below $T_{\rm F}$ [see Fig.~\ref{fig:00}(a)] as an increase in the non-relaxing baseline amplitude, reflecting those components of muon-spin polarization parallel to the local magnetic field which can only be relaxed by dynamic field fluctuations.  

We note that the value of $T_{\mathrm{N}}$ for $x=0$ is considerably suppressed compared to the accepted value of 275~K reported in Ref.~\cite{yamadaB}. We attribute this to oxygen non-stoichiometry, which is an alternative route to doping holes into the CoO$_2$ layers (for La$_{2-x}$Sr$_x$CoO$_{4+\delta}$ the hole density is given by $n_{\rm h}=x+2\delta$), as seen in the cuprate and nickelate systems \cite{wells,tranquadaC}.
Further annealing of an $x=0$ sample led to a restored value of $T_{\rm N}$ of approximately 275~K, however, the lowered amplitude of the $\mu$SR signal and our subsequent NMR measurements indicate that the annealing process introduces impurity phases, particularly at the surfaces of the crystals.
As Sr is introduced within Region I ($x < 0.2$), $T_{\rm N}$ is found to drop abruptly to around 30~K by $x\approx 0.1$. 
This effect, together with that of the oxygen non-stoichiometry, demonstrates the sensitivity of the magnetism within the LSCO system to the concentration of non-magnetic Co$^{3+}$ ions within the CoO$_2$ layers. The presence of holes due to excess oxygen dramatically suppresses the onset temperature compared to the pristine compound, and  $T_{\rm N}$ is further reduced rapidly by further addition of holes via Sr doping.
This is in contrast to both the previously reported phase diagram proposed on the basis of neutron scattering \cite{cwik}, and the predictions of percolation theory for static non-magnetic impurities in two dimensions where long range AFM SO would persist up to $x\approx 0.41$ \cite{newman,vajk}.
However, the effect is much less abrupt than in the superconducting series La$_{2-x}$Sr$_x$CuO$_4$ where long range SO is replaced by an IC spin-glass phase by just $x=0.02$ \cite{matsuda} (we find $\frac{{\rm d} T_{\rm N}}{{\rm d} x} \approx -1000~{\rm K/doped~hole}$ at low $x$; an order of magnitude smaller than in the cuprate series). The freezing temperature $T_{\rm F}$ does not decrease in the same manner, but remains at around 25~K across Region I. The convergence of $T_{\rm N}$ and $T_{\rm F}$ above around $x=0.1$ suggests that for these concentrations the peak in relaxation rate is sensitive to the critical divergence in correlation times accompanying the transition to magnetic SO on the muon timescale. 


{\it Region II:}  As more holes are introduced from further Sr substitution, a marked change in behavior for concentrations $x \geq 0.2$ is encountered, with samples in the region $0.2\leq x\leq 0.5$ showing similar responses, suggestive that they share common features which might relate to stripe ordering.
Stripe order in LSCO has been proposed for $x \geq 0.3$ on the basis of neutron scattering experiments \cite{cwik}. 
Doping away from the parent compound introduces disorder and frustration into the planes and intermediate doping levels lead to short-range stripe correlations which stabilize IC magnetic order. The most robust CO within the LSCO system is checkerboard ordering which occurs at half doping ($x=0.5$) \cite{zaliznyak00,zaliznyak01}, where in-plane CO correlation lengths are largest \cite{cwik}, compared to $x=1/8$ for the cuprate system.
Despite having different electronic properties to the cuprates, LSCO samples with $x=0.33$ and $0.25$ which exhibit disordered stripe CO correlations share the distinctive ``hourglass'' magnetic excitation spectrum \cite{tranquadaD,haydenB,hinkov,lipscombe,xu}, as revealed by inelastic neutron scattering (INS) studies \cite{boothroyd,gaw}.
However, the origin of the hourglass spectrum and the nature of the CO in this region is controversial. 
Due to the insulating nature of the material  and the disparate energy scales of CO and SO \cite{savici} it was suggested that the origin of the IC magnetic order is not stripe-like physics. Furthermore, nanoscale phase separation of regions of the undoped compound and the stable checkerboard CO exhibited by $x=0.5$ doping has been proposed as the source of the hourglass excitation spectrum observed by INS \cite{drees13,drees14}.
Simulations using a cluster spin glass model (CSGM) \cite{andrade} have reproduced the hourglass excitation spectrum \cite{gaw}, where frustrated magnetic ions decorate a background of short-range stripe CO correlations with quenched disorder, strengthening the argument that stripes and hourglass excitation spectrum are intimately linked. 
The results of $\mu$SR and NMR measurements of the $x=1/3$ compound revealed the onset of magnetic order within partially disordered charge and spin stripes at around 35~K, with a further glassy freezing of dynamics involving the slow, collective motion of spins within the magnetic stripes at lower temperatures. 
We find that the $\mu$SR of $x=0.25$ is similar to the $x=0.33$ material, although the slightly broader features preclude the identification of a second freezing feature at lower temperatures.

\begin{figure}[t]
\centering
\includegraphics[width=8cm]{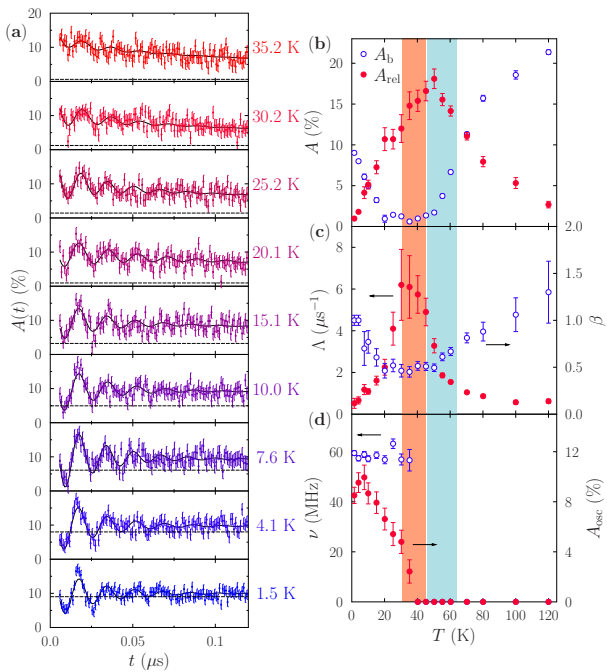}
\caption{\label{fig:20} (a) Asymmetry data for $x=0.2$ measured at S$\mu$S showing  temperature evolution of oscillations, and the non-relaxing baseline (dashed horizontal lines). Solid lines are fits (see text) (b) Amplitude components; (c) longitudinal relaxation rate and exponent $\beta$; (d) amplitude and frequency of the oscillatory component.}
\end{figure}

A key observation is that behavior of $x=0.2$ is clearly different to those samples in Region I, and suggests stripe correlations exist to lower concentrations $x$ than previously believed. Asymmetry spectra obtained at S$\mu$S [Fig.~\ref{fig:20}(a)] show spontaneous oscillations for temperatures below around 40~K, which were best fitted to the single-frequency relaxation function
$A(t) = A_{\rm osc}\cos(2 \pi \nu t + \varphi) {\rm e}^{- \lambda_{\rm osc} t} + A_{\rm rel}{\rm e}^{-( \Lambda t)^{\beta}} + A_{\rm b},$
for the time window $t \leq 4.6~{\rm \mu s}$, where the transverse relaxation rate $\lambda_{\rm osc}$ was fixed to its average value of $39~{\rm \mu s^{-1}}$.
Broad peaks in the temperature dependence of both longitudinal relaxation rate $\Lambda$ and amplitude $A_{\rm rel}$ [Fig.~\ref{fig:20}(b), (c)] 
suggest a freezing temperature $T_{\rm F}=38(7)~{\rm  K}$.  
Upon cooling below 20~K, there is a gradual increase in the non-relaxing contribution $A_{\rm b}$ [Fig.~\ref{fig:20}(a), (b)] which points to a more static field distribution. 
Taken together, these results  indicate that slow dynamics within spatially inhomogeneous disordered stripes of magnetic Co$^{2+}$ ions start to freeze out upon cooling below around 40~K, with regions of static, glassy SO gradually appearing as temperatures are lowered further. This description is consistent with a cluster glass, previously observed only at higher $x$ \cite{gaw}.

The $x=0.5$ compound has been of special interest as it has been found to support checkerboard CO below around $T_{\rm CO} \approx 800~{\rm K}$ \cite{zaliznyak00,zaliznyak01,helme}. Susceptibility and neutron scattering measurements reveal magnetic correlations appearing for $T \lesssim 60~{\rm K}$ \cite{moritomo,helme} and a spin freezing transition at around 30~K dominated by $180^{\circ}$ antiferromagnetic interactions between Co$^{2+}$ ions across the non-magnetic Co$^{3+}$ sites \cite{helme}. Our muon data show a broad drop in $\overline{A}_0$ on cooling through around 30~K, coinciding with a peak in relaxation rate and the appearance of spontaneous (heavily damped) oscillations indicating the onset of long range SO. No features are observed around 60~K, where magnetic moments are still fluctuating outside of the muon timescale.



{\it Region III:} For Sr concentrations  $x\gtrsim 0.5$, neutron measurements have revealed short range IC magnetic correlations on the neutron timescale \cite{sakiyama}, where glassy SO is likely to be accompanied by short-range CO. 
Our measurements in Region III \cite{SIfootnote} detect no long range SO on the muon timescale down to the lowest measured temperatures ($\approx 1.5$~K) for the compounds $x=0.67,~0.75$ and 0.9. For these samples, no muon precession is resolvable and full initial polarization ($\propto A_0$) is maintained across all temperatures. 
Spectra were found to be best fitted to 
$A(t) = A_{\rm f}{\rm e}^{- \lambda_{\rm f} t} +A_{\rm s}{\rm e}^{- \lambda_{\rm s} t} + A_{\rm b}$,
where the initial asymmetry $A_0 = A_{\rm f} +A_{\rm s} + A_{\rm b}$ was fixed to the high-$T$ value of $\overline{A}_0$ for each compound, and the ratio between fast ($\lambda_{\mathrm{f}}$) and slow ($\lambda_{\mathrm{s}}$) relaxation rates was found to be approximately 100.
The fit parameters show qualitatively similar behavior for the three compounds: both the fast component amplitude $A_{\rm f}$ and relaxation rate $\lambda_{\rm f}$ increase as temperature is decreased from around 15~K in a very similar manner to the gradual increase in IC magnetic superstructure Bragg peak intensity observed using neutron scattering for samples with $x \geq 0.6$ \cite{cwik,boothroydA}. We attribute this behavior to the electronic magnetic moments fluctuating more slowly as temperature is lowered. The magnitude of $\lambda_{\rm f}$  as $T$ approaches zero is greatest for $x=0.75$, indicating longer magnetic correlation times for this concentration. The fraction of the total asymmetry contained within the fast relaxing component behaves in the same manner for the three compounds and does not scale with the concentration of non-magnetic Co$^{3+}$ ions, suggesting that there is no phase separation within these samples.
If SO occurs above 1.5~K, then it could be either short-ranged or still fluctuating too rapidly with respect to the muon timescale to be detectable. 


In summary, our $\mu$SR study has elucidated the nature of the magnetic correlations across the phase diagram of the non-cuprate hole-doped layered antiferromagnet \LSCO{} on the muon ($\mu {\rm s}$) time scale, enabling us to identify three distinct regions of behavior. 
For $x \lesssim 0.2$ the ordering temperature $T_{\rm N}$ for the commensurate nnAFM SO is heavily suppressed by the introduction of holes into the CoO$_2$ layers. 
For $0.2 \lesssim x \lesssim 0.6$ ordering temperatures are larger and IC magnetic ordering is likely to be stabilized by stripe correlations in the CO. This region extends to lower hole concentrations than previously thought, implying that the  phase diagram bears a greater similarity to that of La$_{2-x}$Sr$_{x}$NiO$_4$ \cite{ulbrich,hayden,chen,sachan}. Finally, above $x \approx 0.6$ spin fluctuations slow upon cooling, but the system remains paramagnetic down to temperatures of 1.5~K.

\begin{acknowledgements}
Part of this work was performed at S$\mu$S, Paul Scherrer Institut, Villigen, Switzerland and ISIS, Rutherford Appleton Laboratory, UK. We are grateful for the provision of beamtime, and to A. Amato and P. J. Baker for experimental assistance. This research project has been supported by the European Commission 
under the 7\textsuperscript{th} Framework Programme through the `Research Infrastructures' action of the `Capacities' Programme, NMI3-II Grant number 283883. This work is supported by the EPSRC (UK).
\end{acknowledgements}


\begin{thebibliography}{xx}
\bibitem{footnote}
For concentrations $x \geq 0.67$ no drop in initial asymmetry is observed in the data upon cooling. 
The early-time averaging was modified to use the time window $t \leq 0.015~{\rm \mu s}$, and normalized 
between 0 and the maximum value of $\overline{A}_0$.
\bibitem{footnote2}
For this analysis the early-time averaging criterion of $t \leq 0.015~{\rm \mu s}$ was utilized for data measured at S$\mu$S, to take fuller advantage of the increased time resolution relative to ISIS data.
\end{thebibliography}
\end{document}